\documentstyle[amsmath,aps,multicol,prl]{revtex}
\input{epsf}
\begin{document}
\begin{title}
{\bf Statistical mechanics of a discrete nonlinear system }
\end{title}

\author{K.~{\O}. Rasmussen, T. Cretegny\cite{TC}, P.~G. Kevrekidis\cite{PGK}}
\address{Theoretical Division and Center for Nonlinear Studies, 
Los Alamos National Laboratory, Los Alamos, NM 87545}
\author{Niels Gr{\o}nbech-Jensen}
\address{Department of Applied Science, University of California, Davis, CA 95616\\
NERSC, Lawrence Berkeley National Laboratory, Berkeley, CA 94720}
\date{\today}
\maketitle

\begin{abstract}
Statistical mechanics of the discrete nonlinear Schr\"odinger 
equation is studied by means of analytical and numerical techniques.
The lower  bound of the Hamiltonian permits the
construction of standard Gibbsian equilibrium measures for positive
temperatures. Beyond the line of $T=\infty$, we
identify a phase transition, through a discontinuity in the
partition function. The phase transition is demonstrated to manifest itself in 
the creation of 
breather-like localized excitations. Interrelation between
the statistical mechanics and the nonlinear dynamics of the 
system is explored numerically in both regimes.
 \end{abstract}


\begin{multicols}{2}


The pioneering studies of Fermi, Pasta and Ulam \cite{FPU} (FPU) showed that
energy exchange between coupled systems may be suppressed in the presence of
nonlinearity; instead a type of behavior that severely contrasts 
equipartition among the linear
modes is observed. 
The question of whether equipartition of excitation energy always appears
is a contemporary issue in various fields of physics.  Many manifestations of
nonequilibrium and non-equipartition phenomena equivalent to the dynamical
behavior of systems with few degrees of freedom contrasting statistical
mechanics expectations have been observed. Some of these phenomena, and 
therefore
the absence of immediate equipartition expressed in terms of self-trapping of
energy, play an important role for optical storage patterns in nonlinear
fibers, condensed matter physics, and biophysics\cite{thisandthat}. 

A particularity of discrete nonlinear systems
is their ability to sustain strong localization of energy
\cite{ABRY}. This is accomplished via 
 {\em intrinsic localized modes} ({\em breathers})
which are modes that  remain stable for
extremely long times.  So far it is a largely unaddressed problem how to
handle and describe these excitations in a statistical mechanics
framework although it has been argued that breathers may act as virtual
bottlenecks \cite{CRE} delaying the thermalization process.

In this work, we develop a statistical
understanding of the dynamics, including the breathers, in a discrete
nonlinear Schr{\"o}dinger (DNLS) equation.  
The DNLS equation plays a significant role in several branches
nonlinear physics as a simple physical model because
it may approximate many  of the above mentioned nonlinear systems.
We study analytically and numerically the
thermalization of the lattice for $T\geq 0$. We identify the regime in phase
space wherein regular statistical mechanics considerations apply, and
hence, thermalization is observed numerically and explored analytically
using regular, grand-canonical, Gibbsian equilibrium measures.
However, the nonlinear dynamics of the problem renders permissible the
realization of regimes of phase space which would formally correspond to
``negative temperatures''\cite{NegTemp} in the sense of statistical mechanics.
The novel feature of these states is that 
the energy tends to be localized in certain lattice sites 
forming breather-like excitations. Returning to statistical mechanics, such
realizations, which would formally correspond to negative temperatures, are not
possible (since the Hamiltonian is unbounded from above, as is seen by
a simple scaling argument similar to the continuum case 
\cite{LEB}) unless one refines the grand-canonical Gibbsian measure to
correct for that. This correction 
 will necessitate a discontinuity in the partition function
signaling a phase transition that we identify, numerically,
with the appearance of breather modes.

In order to explore and illustrate the scenario described above we 
consider the one-dimensional DNLS equation in the form
\begin{equation}
i\dot \psi_m+ (\psi_{m+1}+\psi_{m-1})+\nu|\psi_m|^2\psi_m=0 \; ,
\label{eqmotion}
\end{equation}
where the overdot denotes time derivative, $m$ is a 
site index, and $\nu$ is a tunable coefficient to the 
nonlinear term\cite{NuRemark}. Equation (\ref{eqmotion}) is the equation of motion,
$\dot \psi_m=-\frac{\partial {\cal H}}{\partial i\psi_m^*},$
where ${\cal H}$ is the Hamiltonian function given by
\begin{equation*}
{\cal H}=\sum_m \left (\psi_m^*\psi_{m+1}+\psi_m\psi_{m+1}^* \right)
+\sum_m\frac{\nu}{2}|\psi_m|^4 \; ,
\end{equation*}
for which $i\psi_m^*$, $\psi_m$ form canonically conjugate pairs of
variables.  In addition to the conserved energy ${\cal H}$, the
quantity ${\cal A}=\sum_m|\psi_m|^2,$ is also conserved by the
dynamics of Eq.\ (\ref{eqmotion}) and serves as the norm of the system.

In order to study the statistical mechanics of the system, we calculate the
classical grand-canonical partition function ${\cal Z}$.
 We first apply the canonical transformation
$\psi_m=\sqrt{A_m}\exp(i\phi_m)$, leading to 
\begin{equation*}
{\cal H} =\sum_m 2\sqrt{A_mA_{m+1}}\cos(\phi_m-\phi_{m+1}) +
\frac{\nu}{2}\sum_mA_m^2 \; .
\end{equation*}

The partition function then becomes,
\begin{equation}
{\cal Z}=\int_0^\infty \int_0 ^{2\pi} \prod_m d\phi_m dA_m
\exp[-\beta({\cal H}+\mu \cal{A})] \; ,
\label{eq:Z}
\end{equation}
where the multiplier $\mu$ is introduced in analogy with a chemical
potential to ensure conservation of ${\cal A}$.  Straightforward
integration over the phase variable $\phi_m$  reduces the symmetrized
partition function to,
\begin{eqnarray*}
\lefteqn{{\cal Z}=(2\pi)^N\int_0^\infty \prod_m dA_m
I_0(2\beta\sqrt{A_mA_{m+1}})\times} \\
&&\exp\left [-\beta\sum_m \left (\frac{\nu}{4}(A_m^2+ A_{m+1}^2)+
\frac{\mu}{2}( A_m +A_{m+1}) \right )\right].
\end{eqnarray*}
This integral can be evaluated exactly in the thermodynamic limit
of a large system ($N\rightarrow \infty$) using the eigenfunctions and eigenvalues
of the transfer integral operator \cite{Krum},
\begin{equation*}
\int_0^\infty dA_{m}\,\kappa(A_m,A_{m+1})\,y(A_m)=\lambda \,y(A_{m+1}),
\end{equation*}
where the kernel $\kappa$ is
\begin{equation}
\label{kernel}
\begin{split}
\kappa(x,z)=&I_0\left(2\beta \sqrt{xz}\right) \times\\
& \exp\left[-\beta\left (\frac{\nu}{4}\left (x^2+z^2\right)
+\frac{\mu}{2} \left ( x+z\right) \right )\right].
\end{split}
\end{equation}
Similar calculations have been performed for the statistical mechanics
of the $\phi^4$ field \cite{Krum}, and for models of DNA denaturation
\cite{DAUX}.  One obtains ${\cal Z}\simeq(2\pi \lambda_0)^N,$
$\mbox{as}~~N\rightarrow \infty$ where $\lambda_0$ is the largest
eigenvalue of the operator. From this expression the usual
thermodynamic quantities such as the free energy, $F$, or specific heat can
be calculated. More importantly, for our purpose we can obtain the
averaged energy density, $h= \langle{\cal H}\rangle/N$,
and the average excitation norm, $a= \langle{\cal A} \rangle/N$ as
\begin{eqnarray*}
 a =-\frac{1}{\beta\lambda_0}\frac{\partial \lambda_0}{\partial \mu},~~
 h=-\frac{1}{\lambda_0}\frac{\partial
\lambda_0}{\partial \beta}-\mu a. 
\end{eqnarray*}

The average excitation norm $a$ can also be
calculated as $a=\frac{1}{{\cal Z}}\int_0^\infty
\prod_m\, dA_m\,A_m \exp\left [ -\beta\left ({\cal H}+\mu{\cal
A}\right ) \right ]$, where the integral again can be calculated using
the transfer technique \cite{Krum} and yields $a
=\int_0^\infty y_0^2(A)A\,dA$, where $y_0$ is the normalized
eigenfunction corresponding to the largest eigenvalue $\lambda_0$ of
the kernel $\kappa$ (Eq.\ (\ref{kernel})).  This shows, that $p(A)=y_0^2(A)$
is the probability distribution function (PDF) for the
amplitudes $A$.

The problem is now reduced to finding the largest eigenvalue
$\lambda_0$ and the corresponding eigenfunction $y_0$ of the transfer
operator, Eq.\ (\ref{kernel}). This we  do numerically. However,
two limits ($\beta\to\infty$ and $\beta\to 0$) are also
amenable to analytical treatment. 

First, notice that the Hamiltonian is bounded from below and one can
 observe that this minimum 
 is realized by a plane wave, $\psi_m=\sqrt{a}\exp im\pi$, whose
energy density is $h=-2a+\frac{\nu}{2} a^2$. 
This relation defines zero temperature, or the $\beta=\infty$ line. 

The high temperature limit is also tractable. When
$\beta \ll 1$ the modified Bessel function in the transfer operator can
be approximated, to leading order, by unity (this amounts to neglecting the
coupling term in the Hamiltonian). This allows us to reduce the
remaining eigenvalue problem to the approximate solution valid for
thermalized independent units,
\begin{equation*}
y_0(A)=\frac{1}{\sqrt\lambda_0} \exp\left [ -\frac{\beta}{4}
\left ( \nu A^2+2\mu A \right ) \right ] \; .
\end{equation*}

Using this approximation we can, enforcing the constraint $\beta \mu=
\gamma$ (where $\gamma$ remains finite as we take the limits $\beta
\to 0$ and $\mu \to \infty$), obtain $h = \nu /\gamma^2$ and $a =
1/\gamma$. Thus, we get $h = \nu a^2$ at $\beta=0$.

\begin{figure}
\epsfxsize=9.cm
\epsffile{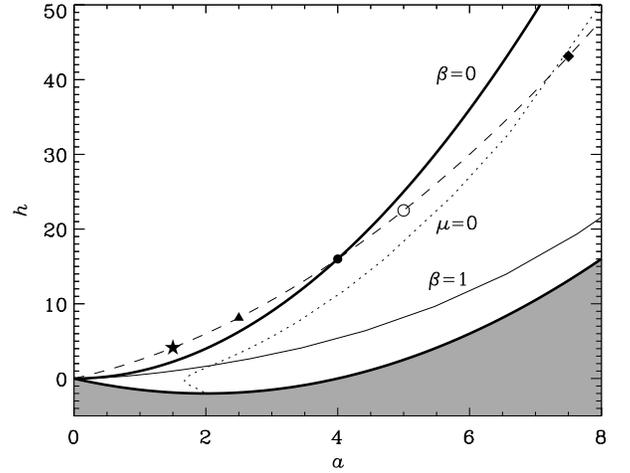}
\caption{Parameter space $(a,h)$, where the shaded area is inaccessible. The
thick lines represents the $\beta=\infty$ and $\beta=0$ lines and thus
bound the Gibbsian regime. The dashed 
line represents the $h=2a+\frac{\nu}{2}a^2$ line along which the
reported numerical simulations are performed (pointed by the symbols).}
\label{fig1}
\end{figure}

Figure 1, depicts (with thick lines) the two parabolas in
$(a,h)$-space corresponding to the $T=0$ and $T=\infty$ limits. Within
this region all considerations of statistical mechanics in the
grand-canonical ensemble are normally applicable and there is a
one-to-one correspondence between $(a,h)$ and $(\beta,\mu)$. Thus,
within this range of parameter space one expects the system to
thermalize in accordance with the Gibbsian formalism. However, the
region of the parameter space that is experimentally (numerically)
accessible is actually wider since it is
possible to initialize the lattice at any energy density $h$ and norm
density $a$ above the $T=0$ line in an infinite system.

A statistical treatment of the remaining domain of parameter space can
be accomplished introducing formally negative temperatures.
But the partition
function (\ref{eq:Z}) is not suited for that purpose since the
constraint expressed in the grand-canonical form fails to bound the
Hamiltonian from above. In all the alternative approaches of the study
of negative temperatures we will have to consider a finite system of
size $N$.  As suggested in \cite{LEB} we can still consider the
grand-canonical ensemble using the modified partition function
${\mathcal{Z}'}(\beta,\mu') = \int \exp(-\beta({\cal H}+ \mu'
{\cal{A}}^2)) \prod_m d{\psi_m}d{\psi^*_m}$, but this introduces long
range coupling and $\mu'$ will have to be of order $1/N$. Now $\beta$
can be negative since ${\cal H}+ \mu' {\cal{A}}^2$ can be seen to be
bounded from above when $\mu' < -\nu/2N$.  The important consequence
of this explicit modification of the measure, is a jump discontinuity
in the partition function, that in turn signals a phase transition.
More explicitly, if one starts in a positive T, thermalizable (in the
Gibbsian sense) state in phase space with $h>0$, and continuously
varies the norm, then one will, inevitably, encounter the
$\beta=0$ parabola.  Hence, in order to proceed in a continuous way,
a discontinuity has to be assigned to the chemical potential.
This discontinuity will destroy the analyticity of the partition
function as the transition line is crossed, and will indicate a
phase transformation according to standard statistical mechanics.

From the microcanonical point of view it is also natural to consider
negative temperatures because it is possible to maximize the energy
under the constraint of fixed norm in a finite system. It can be seen
that the configuration which realizes this maximum is
an exact breather solution, whose total energy and frequency scale as
${\mathcal A}^2$ and $N$, respectively. Thus, the number of microstates
sharing the same energy $E$ will decrease with increasing $E$ if the
norm $\mathcal A$ is kept fixed.  Due to the definition of
temperature ($1/T =\left.\partial S/\partial E\right|_{\mathcal{A}}$),
$T$ becomes negative at high energy density and the $\beta=0$ line is
the line of maximum entropy.  Actually we can say that the constraint
of fixed norm $\mathcal A$ is a ``topological'' reason for large
amplitude breather-like excitations to be expected to appear.

In order to characterize the dynamics of both phases (above and under
the $\beta=0$ line) and to verify that the system does relax to a
thermalized state, we perform numerical experiments.
We restrict the parameters $(a,h)$ to the dashed-line of
Fig.\ \ref{fig1}, choosing a perturbed phonon with
wavevector $q=0$ ($\psi_m=\sqrt{a}$), for which the energy norm
relationship is $h=2a+\frac{\nu}{2}a^2$ as initial condition.
An infinitesimal perturbation
to such a linearly unstable mode for systems of oscillators or
nonlinearly coupled particles, is well-known \cite{CRE,DDP} to give
rise to long-lived localized
excitations via modulational instabilities. For these initial
conditions, the important question is whether the same
phenomenology appears in the DNLS system; i.e., whether relaxation to
equilibrium is really achieved and whether we can observe different
qualitative behavior on the two sides of the $\beta=0$ line.

\begin{figure}
\epsfxsize=9.cm
\epsffile{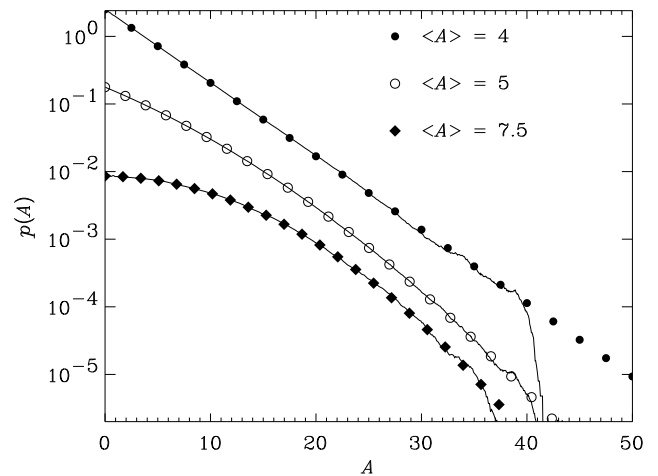}
\caption{Distribution of $A=|\psi|^2$ for three cases under (and on) the
transition line. The solid lines show the results of simulations and
the symbols are given by the transfer operator. Curves are vertically
shifted to facilitate visualization. }
\label{fig2}
\end{figure}

Figure 2 shows three typical examples of what can be observed when
the energy-norm density point lies below the $\beta=0$ line (the symbols
refer to Fig.\ 1). The $q=0$ wave is unstable and the energy density
forms small localized excitations but their lifetime is not very long
and, rapidly, a stationary distribution of the amplitudes $A_m$ is
reached (Fig.\ 2).  Different kinds of initial conditions (with same
energy and norm densities) produced the same results. In conclusion, the
system reaches an equilibrium state which is perfectly recovered by
means of the transfer operator method. Moreover it can be checked on
Fig.\ 2 that the curvature of $\log p(A)$ (i.e., $-\beta$) tends to zero
when $h=\nu a^2$. (The cut-off at high amplitudes is due to finite
size effects). In this domain of parameter space, high amplitude
excitations are highly improbable and can be considered as simple
fluctuations; as shown on Fig.\ \ref{fig2}, large amplitude fluctuations
have been recorded but were checked in the evolution pattern to
disappear rapidly.

The scenario is very different when the energy and norm densities are
above the $\beta=0$ line.  We can observe a rapid creation of
breather excitations due to the modulational instability accompanied
by thermalization of the rest of the lattice. Once created, these
localized excitations remain mostly pinned and because the internal
frequency increases with amplitude their coupling with the small
amplitude radiation is very small. This introduces a new time scale in
the thermalization process necessitating simplectic integration
for as long as $10^6-10^7$ time units in order to reach a stationary
PDF. This can also be
qualitatively justified by the effective long range interactions,
introduced in the modified partition function, which will produce
stronger memory effects as one observes regimes in phase space which
are further away from the transition line (since the long range
interaction will be stronger).

\begin{figure}
\epsfxsize=8.9cm
\epsffile{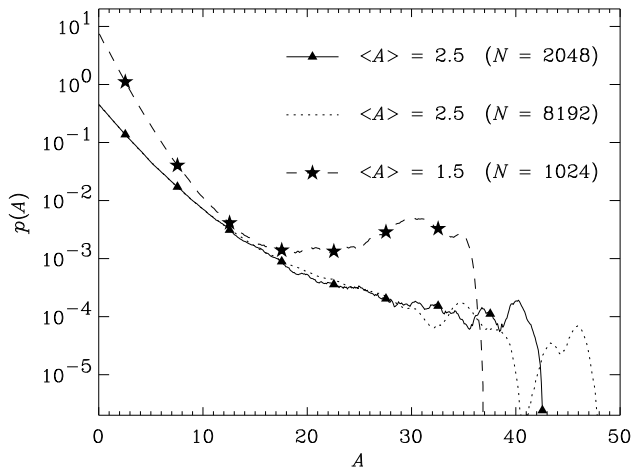}
\caption{Distribution of $A=|\psi|^2$ for parameters $(h,a)$ above the
transition line (triangles and stars as in Fig.\ \ref{fig1}).}
\label{fig3}
\end{figure}

Typical distribution functions of the amplitudes are shown in
Fig.\ \ref{fig3}. The presence of high amplitude excitations is directly
seen here (more straightforwardly we observed standing breathers in
the spatial pattern). The dotted line represents the PDF in the case
where the initial condition is chosen at random, using a larger
system size: we check that the initial condition seems unimportant,
but the system size does influence the amplitude of the
highest breathers. The cut-off value in the very large system limit, as
well as the persistence of a bump in the PDF, is
still an open question, since we have no prediction for the PDF
above the $\beta=0$ line. However the positive curvature of the PDF at small
amplitudes clearly indicates that the system evolves in a regime of
negative temperature and the appearance of the
phase transformation is signified in the dynamics by the appearance of
these strongly localized persistent breathers.  

The actual dynamics in the negative temperature regime is studied 
more closely in Ref. \cite{KATB}.

 
Finally, we can draw an interesting parallel with what has been
known in plasma physics and hydrodynamics for several years\cite{JJR},
where the appearance of localized structures (of vortices in that
case) is also related to a description in terms of negative
temperatures.

In conclusion, studying the DNLS system, we have been able to quantify
and explain, through analytical calculations supported by numerical
computations, the behavior in different regimes of the ($h,a$) phase
space. We have been able to link the regime of thermalization to
the regime where regular statistical mechanics is applicable in
the Gibbsian sense. Further, we have traced the explanation of the
appearance of localized modes in different regimes in phase space to
the need for a modified measure to ensure normalizability, which
therefore necessitates a phase transition leading to these localized modes.
Our numerical simulations strongly
support this theoretical picture, illuminating this novel
quantitative connection between nonlinear dynamics and statistical
mechanics.

The authors gratefully acknowledge discussions with S.\ Aubry, D.\ Cai,
J.\ Farago, R.\ Jordan, M.\ R.\ Samuelsen and J.\ Sethna.
TC and PGK gratefully acknowledge the
warm hospitality of the Center of Nonlinear studies as well as (PGK) 
fellowship support from the ``A.\ S.\ Onasis'' Public Benefit Foundation.  
This work was performed under the auspices of the U.S.\ Department
of Energy and supported by the Director, Office of Advanced Scientific
Computing Research, Division of Mathematical, Information, and
Computational Sciences of the U.S.\ DOE under contract
number DE-AC03-76SF00098.

\end{multicols}


\begin{thebibliography}{10}
\bibitem[*]{TC} Laboratoire de physique de l'ENS-Lyon, CNRS URA 1325, 69364
Lyon, France. New permanent address: LASSP, Cornell University, Ithaca
NY 14853-2501.
\bibitem[**]{PGK} Permanent address: Rutgers University, Department
  of Physics and Astronomy, 136 Frelinghuysen Rd., Piscataway, NJ
  08854-8019, USA
\bibitem{FPU} E. Fermi, J. Pasta, and S. Ulam, Los Alamos Report No. LA-1940, later 
published in Lect. Appl. Math. {\bf 15}, 143 (1974).
\bibitem{thisandthat} See, e.g., Physica D {\bf 68} (1993), special issue on future
directions of nonlinear dynamics in physical and biological systems, edited 
by P.~L. Christiansen, J.~C. Eilbeck, and R.~D. Parmentier.
\bibitem{ABRY} R.~S. MacKay and S. Aubry, Nonlinearity {\bf 7}, 1623 (1994).
\bibitem{CRE} T. Cretegny, T. Dauxois, S. Ruffo, and A. Torcini,
Physica D {\bf 121}, 109 (1998).
\bibitem{NegTemp} Negative temperatures appear as an artifact of applying
the Hamiltonian, which in this system is not a true measure of the energy,
in the statistical formalism.
\bibitem{LEB} J.~L. Lebowitz, H.~A. Rose and E.~R. Speer, J. Stat
  Phys., {\bf 50}, 657 (1988).
\bibitem{NuRemark} In the numerical examples we take $\nu=1$, since a
rescaling of $\psi_m$ allows restriction to this case.
\bibitem{Krum} S. Aubry, J. Chem. Phys. {\bf 62}, 3217 (1975); 
J.~A. Krumhansl and J.~R. Schrieffer, Phys. Rev. B {\bf 11}, 3535 (1975); 
D.~J. Scalapino, M. Sears, and R.~A. Ferrell,  Phys. Rev. B {\bf 9}, 3409 (1972).
\bibitem{DAUX} T. Dauxois, M. Peyrard, and A.~R. Bishop, Physica D {\bf 66}, 35 (1993).
\bibitem{DDP} I. Daumont, T. Dauxois, and M. Peyrard, 
 Nonlinearity, {\bf 10}, 617 (1997).
\bibitem{KATB} K.~{\O}. Rasmussen, S. Aubry, G.P. Tsironis,
A.R. Bishop, (unpublished).
\bibitem{JJR} See, e.g. J. Glenn and D. Montgomery, J. Plasma Physics {\bf 10}, 107 (1973);
              R.~H. Kraichnan and D. Montgomery, Rep. Prog. Phys. {\bf 43}, 547 (1980).


\end{thebibliography}
\end{document}